\documentclass[secnumarabic,amssymb, nobibnotes, aps, prd]{revtex4}

\usepackage{amssymb}
\usepackage{amsmath}
\usepackage{bm}
\usepackage{graphicx,color}
\usepackage{color}

\def\be{\begin{equation}}
\def\ee{\end{equation}}

\def\te{\end{equation}}
\def\bea{\begin{eqnarray}}
\def\nn{\nonumber\\}
\def\tea{\end{eqnarray}}
\begin{document}

\title{An extension of the momentum transfer model to time-dependent pipe turbulence}
\author{Esteban Calzetta}
\email[]{calzetta@df.uba.ar}
\affiliation{Departamento de F\'\i sica, FCEN, Universidad de Buenos Aires, and IFIBA-CONICET, Pabell\'on I, 1428, Buenos Aires, Argentina}
\begin{abstract}
We analyze a possible extension of Gioia and Chakraborty's momentum transfer model of friction in steady turbulent pipe flows (Phys. Rev. Lett. 96, 044502 (2006)) to the case of time and/or space dependent turbulent flows. The end result is an expression for the stress at the wall as the sum of an steady and a dynamic component. The steady part is obtained by using the instantaneous velocity in the expression for the stress at the wall of a stationary flow. The unsteady part is a weighted average over the history of the flow acceleration, with a weighting function similar to that proposed by Vardy and Brown (Journal of Sound and Vibration 259, 1011 (2003); ibid. 270,  233 (2004)), but naturally including the effect of spatial derivatives of the mean flow, as in the Brunone model (B. Brunone et al., J. of Water Resources Planning and Management 236 (2000)).
\end{abstract}

\maketitle

\section{Introduction}
When a fluid flows through a pipe of circular section and radius $R$ it experiences a pressure drop per unit length of pipe $dp/dx=-2\tau/R$, where $\tau$ is the stress at the wall. $\tau$ has units of energy density. For time-independent flows it is commonly parameterized as in the Darcy-Weisbach formula 

\be
\tau\equiv\frac f{8}\rho\:U^2
\label{Darcy}
\te
where $\rho$ is the density of the fluid and $U$ the average flow velocity across a given transverse section of the pipe. The coefficient $f$ in eq. (\ref{Darcy}) is the so-called friction factor \cite{Sch50,MonYag71,Pop00,McZaSm05,YanJos08}.

For a given pipe, the friction factor is a function of Reynolds number 

\be
\mathrm{Re}=\frac{2RU}{\nu}
\label{Reynolds}
\te
where $\nu$ is the kinematic viscosity of the fluid (as distinct from the dynamic viscosity $\mu=\rho\nu$).
$f$ presents three power law-like regimes separated by transition regions. For laminar flows ($\mathrm{Re}<10^3$) we have the Poiseuille Law

\be
f=64\:\mathrm{Re}^{-1}
\label{Pou}
\te
for developed turbulent flows ($10^3<\mathrm{Re}<10^6$) it obeys the Blasius Law 

\be
f=0.3164\:\mathrm{Re}^{-1/4}
\label{Bla}
\te
and for larger values of Reynolds number it converges to an asymptotic value determined by the pipe roughness $\epsilon$ according to the Strickler Law 

\be
f=0.142\:\epsilon^{1/3}
\label{Str}
\te
$\epsilon$ is defined as the characteristic ratio of wall protrusion height to pipe radius.

G. Gioia and P. Chakraborty \cite{GioCha06} have presented a theoretical model whereby the three power law regimes of the steady flow friction factor are easily derived from the Kolmogorov spectrum of homogeneous, isotropic turbulence. See also  \cite{GioBom02,Gol06,GiChBo06,MehPou08,GutGol08,Tran09,Gioi09,Cal09}. We shall refer to this as the momentum transfer model (MTM). The original presentation of the MTM made contact with the Blasius and Strickler (for rough pipes) asymptotic regimes, while the transition from the Blasius to the Poiseuille regimes \cite{GioCha06,Cal09} was discussed in \cite{Cal10}. The connection between wall friction and turbulent spectrum can be used in two directions. If the turbulent spectrum can be found, then it can be used to derive properties of the wall friction. This has been the spirit of the original presentation \cite{GioCha06}. In  \cite{GutGol08} this approach has been used to explore friction in two dimensional flows, and in \cite{Cal10} to investigate drag reduction by polymer additives. 

Alternatively, we may use properties of friction to derive certain features of the turbulent spectrum. For example, in \cite{MehPou08} the scaling properties of the friction coefficient are used to investigate intermittency in the turbulent part of the flow.

In this paper, we aim to combine both approaches to study friction in time-dependent and non-homogeneous turbulent pipe flow. We shall show that a relatively simple generalization of the MTM yields a reasonable expression for the unsteady friction in the somewhat academic case of inhomogeneous, time-independent flows \cite{Saw07}. However, the same generalization fails to describe correctly unsteady friction for time-dependent, homogeneous flows. We shall consider on a phenomenological basis how the MTM should be modified to capture at least some basic elements of unsteady friction in this second case. If the philosophy of the MTM is upheld, this sheds light on the way a time-dependent mean flow excites turbulent fluctuations. 

The subject matter of time dependent turbulent pipe flows has considerable interest, both theoretical and practical \cite{WySt78,DHOJ55,TuRam83,RamTu83,GZMA05}. The basic framework of analysis depends upon the separation of the flow velocity and pressure into a mean component and a turbulent fluctuation. The action of the turbulent fluctuations on the smooth part of the flow is encoded into the stress at the wall by a suitable constitutive relation. To make things simpler, in most cases of interest the turbulent part of the flow may be regarded as incompressible, since characteristic velocities are much lower than the speed of sound. This allows one to bring the extensive lore on friction in incompressible pipe flow to bear on the problem. We shall also assume that the mean part of the flow is parallel. For an incompressible flow this means the mean flow is independent of position along the pipe. We shall relax the incompressibility assumption when discussing inhomogeneous flows.

Another important simplification is available at very high Reynolds number $\mathrm{Re}$. In this case, mean flow may be regarded as essentially one dimensional, and is determined by the equation of state for the fluid and the conservation laws for mass and linear momentum, namely

\be
\frac{\partial\rho}{\partial t}+\frac{\partial\left(\rho U\right)}{\partial z}=0
\label{mass}
\te

\be
\frac{\partial\left(\rho U\right)}{\partial t}+\frac{\partial\left(\rho U^2\right)}{\partial z}=-\frac{\partial p}{\partial z}-\frac {2\tau}R
\label{RANS}
\te
where $\tau$ is the stress at the wall. It is convenient to build in the known friction factor for steady flow by writing  $\tau=\tau_s+\tau_t$. Here $\tau_s$ is the steady part of the friction factor given by eqs. (\ref{Darcy}). $\tau_t$ is the dynamic contribution to friction. 

Most work on time - dependent turbulent pipe flows simply neglect $\tau_t$ \cite{transient}. This approach is sensible at low frequencies, but it is bound to fail eventually \cite{BeSiVi00,CarGun01,ChaLi09}. Indeed that is the case in time-dependent laminar flows, where the time dependent friction may be computed with great accuracy \cite{Zie66}. 

In the literature \cite{VaFi06,ZWL07,AdLe06,AdLe09,Stor10,ZKU11}, it is common to group the different approaches to friction in time-dependent, inhomogeneous flows under two broad headings, namely the convolution method (CM) \cite{Zie66,VarBro03,VarBro04,John11} and the instantaneous acceleration method (IAM) \cite{Bru00}, which in turn may be derived form extended irreversible thermodynamics (EIT) \cite{AGM00}. These approaches are described in next Section. 

To apply the MTM under unsteady conditions we shall follow the same strategy we have used successfully to study drag reduction from polymer additives in \cite{Cal10}. Namely, we shall substitute the turbulent dynamics in the stationary case by a linear stochastic equation which reproduces the Kolmogorov spectrum. Then we shall write down a  generalization of this equation to include a nontrivial background velocity. We shall solve the generalized Langevin equation to compute the spectrum under the new conditions, and thereby derive the friction by following the MTM prescription.

The fact that the correlations of turbulent fluctuations may be cast as stochastic averages over a suitable Langevin dynamics is just a particular application of a  basic theorem in nonequilibrium field theory \cite{GRLE98,Ram07,CalHu08}. The problem is that to really derive this stochastic dynamics from the Schwinger-Dyson equations we need a detailed knowledge of the fluctuation self correlations that is presently unavailable, even for steady flows. Therefore the best we can do is to motivate a particular ansatz for the Langevin equation, by using the few known facts about the spectrum, physical common sense and Occam's razor. 

As a matter of fact, this approach works well for time-independent, inhomogeneous flows. Generally speaking, the MTM understands wall friction in steady flows as the incoherent sum of contributions from turbulent eddies of different sizes. The contribution from each eddy is proportional to its characteristic velocity, which in turn is proportional to the mean flow velocity. A natural generalization of this scenario to inhomogeneous flows is to assume that the eddies contributing to friction at a given position have been created at different points upstream, and then advected by the mean flow (being slowly dissipated in the process)\cite{Krai64,Krai65,BeLv87}. Following this line of argument, it is natural to assume that the relevant eddy velocity is proportional to the mean velocity at the point of creation, rather than at the point where wall friction is being computed. With just this set up we shall be able to arrive at a Brunone-like formula for the unsteady friction.

However, the same approach fails to reproduce the Vardy and Brown CM model of turbulent friction in fully rough flows, which we shall take as paradigmatic, in the case where the mean flow is homogeneous but time dependent. Consider for example the case of an uniformly accelerating flow \cite{AKV10}, and let us call ``now'' the time when we want to compute wall friction. At any time in the past mean velocities are lower than now, and therefore the wall friction computed as above will be necessarily lower than the friction for a steady flow with the present instantaneous mean velocity. Vardy and Brown's CM, and most of the models in the literature, predict just the opposite.  Confronted with this fact we shall seek a minimal modification of the Langevin equation which reproduces at least the basic elements of the CM, without including more free parameters than already present in the Vardy and Brown approach. Although this will not teach us much about unsteady friction, since it is essentially a transcription of the Vardy and Brown model into the Gioia and Chakraborty language, it is potentially revealing concerning the way unsteady pipe flows excite turbulence.

The rest of the paper is organized as follows. In next section we include some necessary background material. We first review the basic approaches to unsteady friction \cite{BeSiVi00,CarGun01}, with emphasis on those of Zielke \cite{Zie66}, Vardy and Brown \cite{VarBro03,VarBro04} and Brunone et al. \cite{Bru00}. Then we introduce the MTM, following closely the original presentation of \cite{GioCha06}. We finally describe the stochastic approach to the MTM previously introduced in \cite{Cal10}. The following two sections are the bulk of the paper, since there we generalize the MTM to time-independent inhomogeneous and homogeneous, time-dependent situations, respectively.  We conclude with a few final remarks.

\section{Basic background}

\subsection{Main approaches to transient friction}
In this subsection we shall briefly review the main approaches to transient friction in the literature.

Zielke \cite{Zie66} provides a thorough discussion of transient friction in laminar flows. By assuming the flow may be regarded as parallel and incompressible, the full set of continuity and Navier-Stokes equations \cite{CarGun01} is reduced to a single equation for a radially dependent mean velocity $u\left[z,r,t\right]$, related to $U\left[z,t\right]$ in eqs. (\ref{mass}) and (\ref{RANS}) through

\be
U\left[z,t\right]=\frac 2{R^2}\int_0^Rrdr\:u\left[z,r,t\right]
\label{zielke1}
\te
If we further neglect $z$-derivatives and nonlinear terms, this single equation reads

\be
\frac{\partial u}{\partial t}-\frac{\nu}r\frac1r\frac{\partial }{\partial r}r\frac{\partial u}{\partial r}=-\frac1{\rho}\frac{\partial p}{\partial z}
\label{zRANS}
\te
Average of this equation over a cross section yields eq. (\ref{RANS}) with the identification

\be
\tau =-\nu\rho \frac{\partial u}{\partial r}\left[z,R,t\right]
\label{ztau}
\te
Because  equation (\ref{zRANS}) is linear and $z$ appears only as a parameter, it admits a solution where

\be
u\left[z,r,t\right]=\frac1{\rho}\int_{-\infty}^tdt'\:g\left[r,t-t'\right]\frac{\partial p}{\partial z}\left[z,t'\right]
\label{zu}
\te
Averaging this equation over a cross section yields a similar relationship

\be
U\left[z,t\right]=\frac1{\rho}\int_{-\infty}^tdt'\:G\left[t-t'\right]\frac{\partial p}{\partial z}\left[z,t'\right]
\label{zu2}
\te
We may consider this as an integral equation for $p$ for a given $U$, invert it and rewrite eq. (\ref{zu}) as

\be
u\left[z,r,t\right]=\int_{-\infty}^tdt'\:g_1\left[r,t-t'\right]U\left[z,t'\right]
\label{zu3}
\te
and finally use this equation and (\ref{ztau}) to obtain a linear relationship between the stress at the wall and the mean velocity.  Thus unsteady friction is naturally expressed as a convolution over the history of the mean flow.

\be
\tau_t\left[z,t\right]=\rho\int_{-\infty}^tdt'\:W\left[t-t'\right]\frac{\partial}{\partial t'}U\left[z,t'\right]
\label{ztt}
\te
The CM attempts to retain this pattern in the turbulent case, which requires some ad hoc assumptions \cite{Pot08}. For example, the Vardy and Brown weighting function is derived by writing a linear equation for the mean velocity along the lines of eq. (\ref{zRANS}), but with an effective viscosity that depends on the radial distance to the pipe. Although more complex, the analysis follows the same steps as Zielke's. The main result of the analysis is the ``weighting function'' $W\left[t\right]$. The particular weighting function derived by Zielke has been found to be an accurate description of transient friction both for laminar and low Reynolds number turbulent flows \cite{BeSiVi00}. Vardy and Brown \cite{VarBro03,VarBro04} and others \cite{others,VSBSL06} have derived weighting functions appropriate for fully developed turbulent flows. Most proposals share many features, like a pronounced peak or else an integrable singularity as $t\to 0$ and an exponential fall off for large $t$. We shall take the Vardy and Brown weighting function as paradigmatic \cite{VarBro04}. The Vardy-Brown weighting function is best introduced through its Laplace transform

\be
W_{VB}\left[s\right]=\frac {A\sqrt{\pi\nu}}{\sqrt{s+\left(C\omega\right)}}
\label{VBL}
\te
which in time domain becomes

\be
W_{VB}\left[t\right]={A}\sqrt{\frac{\nu}t }{e^{-C\omega t}}
\label{VB}
\te
where $\omega$ is the viscous dissipation frequency

\be
\omega=\frac{\nu }{R^2}
\label{omegavisc}
\te
The constants $A$ and $C$ depend on Reynolds number and roughness. For fully rough flow they are given by \cite{VarBro04}

\be
A=0.0206\:\sqrt{{\mathbf{Re}}}\:\left(\frac{\epsilon}2\right)^{0.39}
\label{vba}
\te

\be
C=0.352\:{\mathbf{Re}}\:\left(\frac{\epsilon}2\right)^{0.41}
\label{vbc}
\te
Observe that ${\mathbf{Re}}$ in eqs. (\ref{vba}) and (\ref{vbc}) is not the instantaneous Reynolds number eq. (\ref{Reynolds}), but rather a fiducial Reynolds number, e. g., the Reynolds number of a preexisting steady flow \cite{VarBro04}. By construction, ${\mathbf{Re}}$ is space and time independent.

While the CM represents unsteady friction as an average over the history of the mean flow at a fixed point, the IAM builds unsteady friction from time and space derivatives of the mean flow at a given time. The Brunone ansatz for the transient friction reads \cite{Bru00} 

\be
\tau_t\left[z,t\right]=k_B\rho R \left\{\frac{\partial U}{\partial t}-a\frac{\partial U}{\partial z}\right\}\left[z,t\right]
\label{brutt}
\te
Brunone et al.\cite{Bru00}  identify $a$ with the speed of the pressure wave, while the EIT derivation of eq. (\ref{brutt}) suggests it should be chosen as the mean velocity $U$ itself \cite{AGM00}. In practice, both $k_B$ and $a$ are often regarded as a free parameters \cite{SzMi07,StNi09}. 

The signs in eq. (\ref{brutt}) are also a matter of some controversy \cite{LVSB01,VBSL06b,John11}. Among several proposals we mention \cite{BeSiVi00}

\be
\tau_t\left[z,t\right]=k_B\rho R \left\{\mathrm{sign}\left[U\right]\frac{\partial U}{\partial t}+a\left|\frac{\partial U}{\partial z}\right|\right\}\left[z,t\right]
\label{bvtt}
\te
The Brunone constant $k_B$ may be derived from eqs. (\ref{ztt}) and (\ref{VB}) if we require that CM and IAM agree for a $z$-independent flow with constant acceleration

\be
k_B=\frac1R\int\:dt\:W\left[t\right]=A\sqrt{\frac{\pi}C}=0.06\left(\frac{\epsilon}2\right)^{0.185}
\label{kb}
\te

\subsection{The original momentum transfer model}

Let us consider a stationary flow within a straight pipe of circular section and radius $R$. Let $z$ be the coordinate along the pipe. The  flow may be decomposed into mean flow  and fluctuations as $\mathbf{U}^p=U\hat{z}^p+\mathbf{u}^p$, where $\hat{z}^p$ is the unit vector in the $z$ direction. $U$ is defined by the condition that $A\left\langle \rho\right\rangle U$ gives the flux across the section of the pipe, where $A$ is the area of the cross section and $\left\langle \right\rangle$ is the average over the section. Henceforth we shall drop the angle brackets on $\left\langle \rho\right\rangle$

In the central region, each scale $\delta$ is associated to a velocity $u\left[\delta, U\right]$. The MTM claims that
the stress at the wall is given by

\be
\tau=\rho U u\left[\delta^*R,U\right]
\label{8}
\te
$\delta^*R$ is a characteristic length which is equal to the Kolmogorov scale for flow in smooth pipes (leading to the Blasius Law) or else to the pipe roughness $\epsilon R$ for rough pipes at large Reynolds number, leading to the Strickler Law \cite{GioCha06}. We shall assume the latter in what follows. For the recovery of the Poiseuille regime see \cite{Cal10}.

The velocity $u\left[\delta^*R,U\right]$ which appears in eq. (\ref{8})  is defined from the mode decomposition of the turbulent energy

\be
u\left[\epsilon R,U\right]^2\equiv\left\langle \mathbf{u}^2\left[\epsilon R,U\right]\right\rangle=2\int_{\left(\epsilon R\right)^{-1}}^{\infty}\:dk\:E\left[k\right]
\label{global2}
\te
Of course to apply this formula, we need to know the spectrum of turbulent fluctuations for turbulent pipe flow. Gioia and Chakraborty get around this difficulty by assuming that the actual spectrum, at least in the relevant short wavelength sector, is the same as for an unbounded homogeneous flow, namely the Kolmogorov spectrum $E\left[k\right]=C_K\zeta^{2/3}k^{-5/3}$, where $C_K\approx 1.5$ is the so-called Kolmogorov constant \cite{MCCO94}. (Although this simple picture must be modified in the dissipative and energy ranges, these corrections are not relevant at large Reynolds numbers and we shall not discuss them explicitly; \cite{GioCha06} works with a more realistic spectrum, see also \cite{Cal09}). Even after this drastic simplification, a crucial input is still missing, namely the value of  $\zeta$, which is  the energy flux feeding the Richardson cascade. $\zeta$ is usually written in terms of a turbulent velocity scale $u_T$ as

\be
\zeta=\frac{u_T^{3}}{R}
\label{302}
\te
For $u_T$, Gioia and Chakraborty assume a linear dependence on the mean velocity

\be
u_T=\kappa U
\label{xx}
\te
It is interesting to contrast eq. (\ref{xx}), for example, with the usual expression for the turbulent velocity scale in terms of a mixing length $L$, namely $u_T\approx L dU/dr$ \cite{Sch50,Pop00}. The goal of the mixing length approach is to obtain the turbulent velocity scale from local properties of the mean flow. An unbounded homogeneous mean flow would excite no turbulence, out of galilean invariance, and so mean flow derivatives must be called forth. To the contrary, eq. (\ref{xx}) is not intended as a local relationship, but as a global one. In particular, the boundary condition at the wall plays an essential, if implicit, role. If we could solve the full problem of turbulent pipe flow, we expect to find (if only out of dimensional reasons) something like eq. (\ref{xx}) at least close enough to the pipe axis \cite{LLPP05,LPPZ06,LPR08,ESHW07,ZhGh06,RNR09}. Because we cannot truly derive eq. (\ref{xx}), we cannot estimate the value of the parameter $\kappa$. However, if we leave it as a free parameter we may compute 

\be
u\left[\epsilon R,U\right]^2=3C_K\kappa^2\epsilon^{2/3}U^2
\label{str1}
\te
then eq. (\ref{8}) reduces to eq. (\ref{Darcy}), provided we identify

\be
f=8\sqrt{3C_K}\kappa\epsilon^{1/3}
\label{str2b}
\te
Comparing with eq. (\ref{Str}) we obtain  $\kappa=0.008$. We see that even in its original formulation the MTM is not free from ad hoc assumptions.

\subsection{Stochastic approach to the MTM}

It is clear that the full Navier-Stokes equation is too complex for analysis, unless numerically. To make progress, we shall substitute the Navier-Stokes equation by a linear stochastic one, devised to give the right spectrum of turbulent fluctuations, and therefore enabling us to compute $u\left[\epsilon R,U\right]$. 

Let us begin by Fourier decomposing the fluid velocity

\be
\mathbf{u}^p\left(\mathbf{x},t\right)=\int\:\frac{d\mathbf{k}}{\left(2\pi\right)^3}\:e^{i\mathbf{k}\mathbf{x}}\mathbf{u}^p_{\mathbf{k}}\left[t\right]
\label{Fourier}
\te
We may think of the Fourier components $\mathbf{u}^p_{\mathbf{k}}\left[t\right]$ as the velocity associated to eddies of size $k^{-1}$ at time $t$. We conceive the dynamics of these velocities as the balance of two processes: on one hand they draw energy and momentum from larger eddies (and ultimately the mean flow), on the other they are subject to dissipation from interaction with smaller eddies \cite{Hei48,Cha49}. We model the first process by a stochastic driving force, the second one by an eddy viscosity. In steady homogeneous turbulence both processes balance each other, much like in the fluctuation-dissipation theorem of linear nonequilibrium thermodynamics \cite{McCKiy05}.

We postulate for the Fourier components a dynamic equation

\be
\left[\frac{\partial}{\partial t}+\sigma_k\right]\mathbf{u}^p_{\mathbf{k}}\left[t\right]=F^p_{\mathbf{k}}\left[t\right]
\label{dynlinearhom}
\te
This means that at any time $t$ the velocity of eddies of size $k^{-1}$ undergoes a random increment $F^p_{\mathbf{k}}dt$ over a lapse $dt$, which subsequently decays exponentially with a mean lifetime $\sigma_k^{-1}$. The velocity increments are Gaussian, and uncorrelated if we look at eddies of different sizes and/or at different times

\be
\left\langle F^p_{\mathbf{k}}\left[t\right]F^q_{\mathbf{k'}}\left[t'\right]\right\rangle=\left(2\pi\right)^5\delta\left(\mathbf{k}+\mathbf{k'}\right)\delta\left(t-t'\right)\Delta^{pq}_{\mathbf{k}}N_k
\label{dynselfbhom}
\te
$\Delta^{pq}_{\mathbf{k}}$ is a projector that enforces incompressibility

\be
\Delta^{pq}_{\mathbf{k}}=\delta^{pq}-\frac{\mathbf{k}^p\mathbf{k}^q}{k^2}
\label{delta}
\te
A representation like this may be derived from the functional approach to turbulence, where the left hand side of eq. (\ref{dynlinearhom}) is identified as the inverse retarded propagator, and the self-correlation eq. (\ref{dynselfbhom}) is given by a self-energy\cite{MCCO94,Cal09b}. We shall be content to propose simple expressions for $\sigma_k$ and $\mathbf{N}_k$ to reproduce the known turbulent spectrum.

In the inertial range, we expect $\sigma_k$ and $\mathbf{N}_k$ to depend on the only dimensionful parameter $\zeta$ from eq. (\ref{302}). Let us assume the linear ansatz eq. (\ref{xx}). 
On dimensional grounds $\sigma_k\propto\left(k^2\zeta\right)^{1/3}$ \cite{YII02}. An important insight from ref. \cite{Cal10} is that to obtain the Virk Asymptote for drag reduction by polymer additives, it is necessary to assume that $\sigma_k$ is inversely proportional to Reynolds number 

\be
\sigma_k=\frac{\lambda}{\kappa\mathrm{Re}}\left(k^2\zeta\right)^{1/3}
\label{nucero}
\te
where $\lambda$ is a pure number. An important consequence of eq. (\ref{nucero}) is that $\sigma_k$ is actually independent of the local mean velocity $U$

\be
\sigma_k=\frac {\lambda}2\omega\left( Rk\right)^{2/3}
\label{drag}
\te
where $\omega$ is defined in eq. (\ref{omegavisc}).
This will entail an important simplification in what follows.

On the other hand, solving the Langevin eq. (\ref{dynlinearhom}) gives the ``fluctuation dissipation theorem'' $k^2N_k=E\left[k\right]\sigma_k$ and therefore

\be
N_k=C_K\sigma_k\zeta^{2/3}k^{-11/3}=\frac12C_K\lambda\kappa^2\omega\frac{U^2}{k^3}
\label{str3}
\te
Observe that $N_k\propto U^2$ or else $F^p_{\mathbf{k}}\propto U$. This is the analog to eq. (\ref{xx}) in the stochastic approach. We shall see presently that this relation must be generalized for time-dependent flows.

This concludes the review of the stationary case.

\section{From MTM to IAM}

We now face the problem of generalizing the above effective linear model to the case where the background velocity $U$ is position dependent, though time independent \cite{Saw07}. We cannot reduce this problem to the homogeneous one, for example, by considering turbulence in the presence of a mean velocity gradient \cite{Div11}, because, as we already remarked, the relationship of the turbulent fluctuations to mean velocity is not purely local. Therefore we shall proceed by developing a natural, but to some extent arbitrary, generalization of the stochastic approach to the MTM.

The basic idea is that at any time, and within the volume of fluid between two cross sections at positions $z$ and $z+\epsilon R$, the flow remains traslation invariant enough that we may still introduce a Fourier transform, though the Fourier amplitudes now carry a ``slow'' variable indicating where the flow is being Fourier analyzed. In formulae

\be
\mathbf{u}^p\left(\mathbf{x},t\right)=\int_{Rk>\epsilon^{-1}}\:\frac{d\mathbf{k}}{\left(2\pi\right)^3}\:e^{i\mathbf{k}\mathbf{x}}\mathbf{u}^p_{\mathbf{k}}\left[t;z\right]
\label{dynFourier}
\te
We could make this definition precise by introducing centroid and relative variables \cite{EdMcC72}, but an intuitive idea is sufficient for present needs. See also \cite{Cal09b,CalHu08}. Please observe that we shall never need to compute other than instantaneous velocity correlations between two points at a distance less than $\epsilon R$ apart.

To obtain a dynamics for the Fourier amplitudes in eq. (\ref{dynFourier}) we assume that, besides the dissipation and noise already accounted for in eq. (\ref{dynlinearhom}), turbulence is advected by the mean flow \cite{Krai64,Krai65,BeLv87}. Therefore we write

\be
\left[\frac{\partial}{\partial t}+U\left(z,t\right)\frac{\partial}{\partial z}+\sigma_k\right]\mathbf{u}^p_{\mathbf{k}}\left[t;z\right]=F^p_{\mathbf{k}}\left[t;z\right]
\label{dynlinear}
\te
When the Langevin equation is systematically derived from the Schwinger-Dyson equations, it is seen that the left hand side of eq. (\ref{dynlinear}) is the inverse of the causal propagator for linearized fluctuations in the turbulent flow. The evolution of these fluctuations is determined mostly by their interaction with the turbulent eddies and it is robust with respect to changes in the external conditions. Therefore it makes sense to
assume that the mean lifetime of a fluctuation is the same as in a fiducial steady flow. Then $\sigma_k$ is still given by eq. (\ref{drag}), and it is position and time-independent. 
The solution to eq. (\ref{dynlinear}) is

\be
\mathbf{u}^p_{\mathbf{k}}\left[t;z\right]=\int^t_{-\infty}dt'\:e^{-\sigma_k\left(t-t'\right)}F^p_{\mathbf{k}}\left[t';\xi\left[z,t;t'\right]\right]
\label{dynsol}
\te
where $\xi$ obeys

\be
\left[\frac{\partial}{\partial t}+U\left(z,t\right)\frac{\partial}{\partial z}\right]\xi\left[z,t;t'\right]=0
\label{Lag}
\te
with boundary condition

\be
\xi\left[z,t;t\right]=z
\label{Lag2}
\te
To see the meaning of $\xi$, observe that we could define a function $z_0\left(t\right)$ by holding $t'$ and $\xi\left[z_0\left(t\right),t;t'\right]=\xi_0$ constant, and then we get $dz_0/dt=U\left[z_0\left(t\right),t\right]$ and $z_0\left(t'\right)=\xi_0$, so $\xi$ is a Lagrangian coordinate for the particles, with respect to the mean flow.

We now generalize the MTM ansatz eq. (\ref{8}) to inhomogeneous situations as

\be
\tau\left(z,t\right)=\rho\left(z,t\right) U\left(z,t\right) u\left(z,t\right)
\label{dyn8}
\te
where

\be
u\left(z,t\right)^2=\int_{Rk>\epsilon^{-1}}\:\frac{d\mathbf{k}}{\left(2\pi\right)^3}\int_{Rk'>\epsilon^{-1}}\:\frac{d\mathbf{k}'}{\left(2\pi\right)^3}\:e^{i\left(\mathbf{k}+\mathbf{k'}\right)\mathbf{x}}\left\langle \mathbf{u}^p_{\mathbf{k}}\left[t;z\right]\mathbf{u}^p_{\mathbf{k'}}\left[t;z\right]\right\rangle
\label{dynu}
\te
where $\mathbf{x}$ is any point close enough to the cross section through $z$. To compute the expectation value we need to make some assumption regarding the noise self-correlation. We assume the random accelerations $F^p_{\mathbf{k}}\left[t;z\right]$ are still Gaussian and uncorrelated at different scales and/or times, and that they scale  as the local mean flow velocity. This leads to

\be
\left\langle F^p_{\mathbf{k}}\left[t;z\right]F^q_{\mathbf{k'}}\left[t';z'\right]\right\rangle=\left(2\pi\right)^5\delta\left(\mathbf{k}+\mathbf{k'}\right)\delta\left(t-t'\right)\Delta^{pq}_{\mathbf{k}}\frac{C_K\lambda\kappa^2\omega}{2}{U}\left(t,z\right){U}\left(t,z'\right)
k^{-3}
\label{dynselfb}
\te
where we have build in the requirement that it reduces to eq. (\ref{dynselfb}) if the mean velocity is homogeneous.
Using eqs. (\ref{dynselfb}),  (\ref{dynsol}), (\ref{drag}) and (\ref{str2b}) we get

\be
u\left(z,t\right)^2=\frac 23\left(\frac f8\right)^2\int_{x>1}\:\frac{dx}{x}\int^t_{-\infty}\Omega dt'\:e^{-\Omega\left(t-t'\right)x^{2/3}}{U}^2\left[\xi\left[z,t;t'\right],t'\right]
\label{dynu2b}
\te
where 

\be
x=\epsilon Rk
\te
and 

\be
\Omega=\lambda\epsilon^{-2/3}\omega.
\label{cut}
\te
Observe that

\be
\frac 23\int_{x>1}\:\frac{dx}{x}\int^t_{-\infty}\Omega dt'\:e^{-\Omega\left(t-t'\right)x^{2/3}}=1
\label{ints}
\te
Therefore, for steady flow we recover eqs. (\ref{Darcy}) and (\ref{Str}). In the general case, we have $\tau=\tau_s+\tau_t$. Therefore, if $\tau =\rho U u$ and $\tau_s=\left(f/8\right)\rho U^2$, then

\be
u=\frac1{\rho U}\left[\tau_t+\frac f8\rho U^2\right]
\label{ub}
\te
Taking the square of this equation and equating to eq. (\ref{dynu2b}) we get

\be
\frac{\tau_t}{\rho}+\frac{\tau_t^2}{2\rho\tau_s}=\frac 13\left(\frac f{8}\right)\int_{x>1}\:\frac{dx}{x}\int^t_{-\infty}\Omega dt'\:e^{-\Omega\left(t-t'\right)x^{2/3}}\left[{U}^2\left[\xi\left[z,t;t'\right],t'\right]-U^2\left[z,t\right]\right]
\label{dynu3b}
\te
We consider a situation where the mean flow is only weakly inhomogeneous, so the second term in the left hand side of eq. (\ref{dynu3b}) is negligible, and also time independent. Since $z$ is also $\xi\left[z,t;t\right]$, we may approximate

\be
{U}^2\left[\xi\left[z,t;t'\right],t'\right]-U^2\left[z,t\right]=-\left(t-t'\right)2U\frac{dU}{dz}\left.\frac{\partial\xi\left[z,t;t'\right]}{\partial t'}\right|_{t'=t}
\label{manyder}
\te
where

\be
\left.\frac{\partial\xi\left[z,t;t'\right]}{\partial t'}\right|_{t'=t}=-\left.\frac{\partial\xi\left[z,t;t'\right]}{\partial t}\right|_{t'=t}=U\left(z,t\right)\left.\frac{\partial\xi\left[z,t;t'\right]}{\partial z}\right|_{t'=t}=U\left(z,t\right)
\label{toomanyders}
\te
so

\be
\frac{\tau_t}{\rho}=-\left(\frac f{16}\right)\frac{U^2}{\Omega}\frac{dU}{dz}
\label{mtmiam}
\te
We recover the original Brunone formula eq. (\ref{brutt}) specialized to time-independent flows,  provided we identify

\be
ak_B=\left(\frac f{16}\right)\frac{U^2}{R\Omega}
\label{ident}
\te
\section{From MTM to CM}
We now consider the complementary problem where the mean velocity is homogeneous in space but time-dependent. As in the previous Section, there is no simple way of reducing this problem to the homogeneous one. For example, to look at the flow from a comoving, non inertial frame \cite{Spe88} would be of no avail, because it changes the boundary condition at the wall in an essential way. On the other hand, we shall see presently that the ansatz eq. (\ref{dynselfb}) for the noise self-correlation leads to an untenable prediction for transient friction. If we believe in the MTM, we must conclude that a time-dependent mean field excites turbulence in an essentially different way that a time-independent one.

Concretely, in the model we have considered so far, the turbulent velocity at a given scale and time is the sum of velocity increments occurred at all earlier times and upstream positions. Once produced, each elementary increment is exponentially damped while it is advected by the mean flow.

Suppose we had a spatially homogeneous accelerating flow. Let us look at the turbulent velocity at some time $t_{now}$, when the mean flow velocity is $U_{now}$. If the random velocity increments scale as the instantaneous velocity at the time and position of production, as we have assumed so far, then the velocity increments in the past have been smaller than for a steady flow with velocity $U_{now}$. Add the exponential damping to this, and the conclusion is that the turbulent velocity, and so the wall friction according to eq. (\ref{dyn8}), will be less for the accelerating flow than for the steady flow with velocity $U_{now}$. This is contrary to Vardy and Brown CM's and most other models in the literature. 

To save the MTM we must conclude that the random velocity increments in the right hand side of eq. (\ref{dynlinear}) are sensitive not only to the local mean velocity, but to mean velocity derivatives as well. Of course, lacking a full solution for pipe turbulence in time-dependent conditions, our only guidance is phenomenology. On the other hand, not to sacrifice too much predictive power, we would like to build a model with no more free parameters than, say, the Vardy-Brown CM. The Vardy-Brown weighting function has two adjustable parameters, the $A$ and $C$ constants eqs. (\ref{vba}) and (\ref{vbc}). Our model already has one parameter over and above the steady state model, namely the constant $\lambda$ in eq. (\ref{drag}), so we only have room for one new parameter.

The simplest hypothesis is that the local random velocity increments scale not as the instantaneous velocity $U\left[z,t\right]$ but rather as some effective velocity scale $\tilde{U}\left[z,t\right]$ which depends also on the local acceleration. Repeating the arguments in the previous Section we arrive at

\be
\frac{\tau_t}{\rho}+\frac{\tau_t^2}{2\rho\tau_s}=\frac 13\left(\frac f{8}\right)\int_{x>1}\:\frac{dx}{x}\int^t_{-\infty}\Omega dt'\:e^{-\Omega\left(t-t'\right)x^{2/3}}\left[\tilde{U}^2\left[t',x\right]-U^2\left[t\right]\right]
\label{dynu3}
\te
We impose the requirement that for a weakly time-dependent and space independent flow, eq. (\ref{dynu3}), after linearization in $\tau_t$ and $\partial U/\partial t$, should reduce to eq. (\ref{ztt}) with a weighting function consistent with $W_{VB}$ as given in eqs. (\ref{VBL}) and (\ref{VB}). 
When $\tau_t$ is small, we may write

\be
{\tau_t}=\tau_t^{(1)}+\tau_t^{(2)}
\te
where

\bea
\frac{\tau_t^{(1)}}{\rho}&=&-\frac 13\left(\frac f{8}\right)\int_{x>1}\:\frac{dx}{x}\int^t_{-\infty}\Omega dt'\:e^{-\Omega\left(t-t'\right)x^{2/3}}\left[{U}^2\left[t\right]-U^2\left[t'\right]\right]\nn
&=&-\frac 13\left(\frac f{8}\right)\int_{x>1}\:\frac{dx}{x}\int^t_{-\infty}\Omega dt'\:e^{-\Omega\left(t-t'\right)x^{2/3}}\left[\frac{2U\left[t'\right]}{\Omega x^{2/3}}\right]\frac{dU}{dt'}
\label{dynu3a}
\tea

\be
\frac{\tau_t^{(2)}}{\rho}=\frac 13\left(\frac f{8}\right)\int_{x>1}\:\frac{dx}{x}\int^t_{-\infty}\Omega dt'\:e^{-\Omega\left(t-t'\right)x^{2/3}}\left[\tilde{U}^2\left[t',x\right]-U^2\left[t'\right]\right]
\label{dynu3c}
\te
It is clear that if we choose $\tilde{U}=U$ then $\tau_t^{(2)}=0$ and we get an expression for $\tau_t$ with the wrong sign, as expected. Therefore we must contemplate a more general ansatz. 
To keep the number of new parameters to a minimum, we are satisfied with proposing a simple scale-free form

\be
\tilde{U}\left[t,x\right]=U\left[t\right]+T\left(Rk\right)^{\alpha}\frac{\partial U}{\partial t}\left[t\right]
\label{tildeU}
\te
$T$ is a characteristic time to be determined, and it is the single new parameter we shall allow ourselves. After linearization we get indeed eq. (\ref{ztt}) with a weighting function

\be
W\left[t\right]=\frac 23\left(\frac {f}{8}\right){\mathbf{Re}}\:R\omega \:\mathrm{sign}\left[U\right]\int_{x>1}\:\frac{dx}{x} \:e^{- x^{2/3}\Omega t}\left[\frac{\Omega T}{\epsilon^{\alpha}}x^{\alpha}-x^{-2/3}\right]
\label{weight}
\te
To obtain $W\left[t\right]\approx t^{-1/2}$ when $t\to 0$ requires $\alpha=1/3$. Integrating over time we obtain the Brunone constant

\be
k_B=\left(\frac {f}{4}\right)\frac{{\mathbf{Re}}\:\omega }{\Omega }\left[\frac{\Omega T}{\epsilon^{1/3}}-\frac14\right]
\label{ident2}
\te
and then, writing $x^{1/3}=y/\sqrt{\Omega t}$ 

\be
W\left[t\right]=\left(\frac {f}{8}\right){\mathbf{Re}}\:R\omega \:\mathrm{sign}\left[U\right]\left[\frac{\Omega T}{\epsilon^{1/3}}\sqrt{\frac{\pi}{\Omega t}}\:\mathrm{erfc}\left[\sqrt{\Omega t}\right]+\Omega t\:\mathrm{Ei}\left[-\Omega t\right]\right]
\label{Lap2}
\te
where

\be
\mathrm{erfc}\left[x\right]=\frac 2{\sqrt{\pi}}\int_x^{\infty}dy\:e^{-y^2}
\label{erfc}
\te

\be
\mathrm{Ei}\left[x\right]=-\int_{-x}^{\infty}\frac{dy}y\:e^{-y}
\label{ei}
\te
This weighting function decays as $\exp\left(-\Omega t\right)$. To match the Vardy-Brown weighting function eq. (\ref{VB}) we ask

\be
\Omega =C\omega
\label{freq}
\te
where $C$ is defined in eq. (\ref{vbc}). This may be used to fix the constant $\lambda$ in eqs. (\ref{nucero}) and (\ref{drag}), namely

\be
\lambda = C\epsilon^{2/3}
\label{lambda2}
\te
Observe that if we adopt eq. (\ref{lambda2}) then we move beyond the framework of \cite{Cal10}, because there $\lambda$ is supposed not to depend on Reynolds number. It should be observed that ref. \cite{Cal10} deals with relatively low Reynolds numbers, consistent with Blasius' Law in absence of the polymer, while here we are considering very high ones to which Strickler's scaling applies. 

Using eqs. (\ref{Str}), (\ref{vbc}), (\ref{kb}) and (\ref{freq}) into (\ref{ident2}) we get

\be
\frac{\Omega T}{\epsilon^{1/3}}-\frac14=\frac U{8a}=0.47\:\left(\frac{\epsilon}2\right)^{0.26}
\label{ident4}
\te
For example, if $\epsilon /2=10^{-2}$, then the right hand side of eq. (\ref{ident4}) gives $0.15$, and becomes smaller by a factor of $10$ if $\epsilon /2=10^{-6}$. For these values of ${\Omega T}/\epsilon^{1/3}$ the MTM weighting function eq. (\ref{Lap2}) changes sign for relatively low values of $\Omega t$. Nevertheless, the scale-free ansatz eq. (\ref{tildeU}) captures the short time behavior and the overall integral of the weighting function, so it may be trusted to give right results both for strongly accelerating flows and slowly varying flows. In any case, all CM models have only a finite time duration of validity \cite{VTBCSL08a,VTBCSL08b,HAV08}, so we cannot truly say that such a sign reversal is ruled out by experiment \cite{AHV10}.

For larger values of ${\Omega T}/\epsilon^{1/3}$ the sign reversal occurs too late to be of any relevance. This follows from the asymptotic expansions \cite{abramo}

\be
\mathrm{erfc}\left[x\right]\approx\frac {e^{-x^2}}{\sqrt{\pi}x}
\label{erfcas}
\te

\be
\mathrm{Ei}\left[-x\right]\approx -\frac {e^{-x}}{x}
\label{eias}
\te

After the identifications eqs. (\ref{ident2}) and (\ref{freq}), the Vardy and Brown weighting function eq. (\ref{VB}) reads

\be
W_{VB}\left[t\right]=\left(\frac {f}{4}\right){\mathbf{Re}}\:R\omega \left[\frac{\Omega T}{\epsilon^{1/3}}-\frac14\right]\frac{e^{-\Omega t}}{\sqrt{\pi\Omega t}}
\label{VB2}
\te

In fig. 1 we show a plot of both weighting functions as a function of $\Omega t$, divided by  $\left(f/4\right){\mathbf{Re}}\:R\omega $, for ${\Omega T}/\epsilon^{1/3}=.4$. Fig. 2 is a close up of the short $t$ behavior, for which we have chosen a log-log scale.

\begin{figure}[htp]
\centering
\includegraphics[scale=.8]{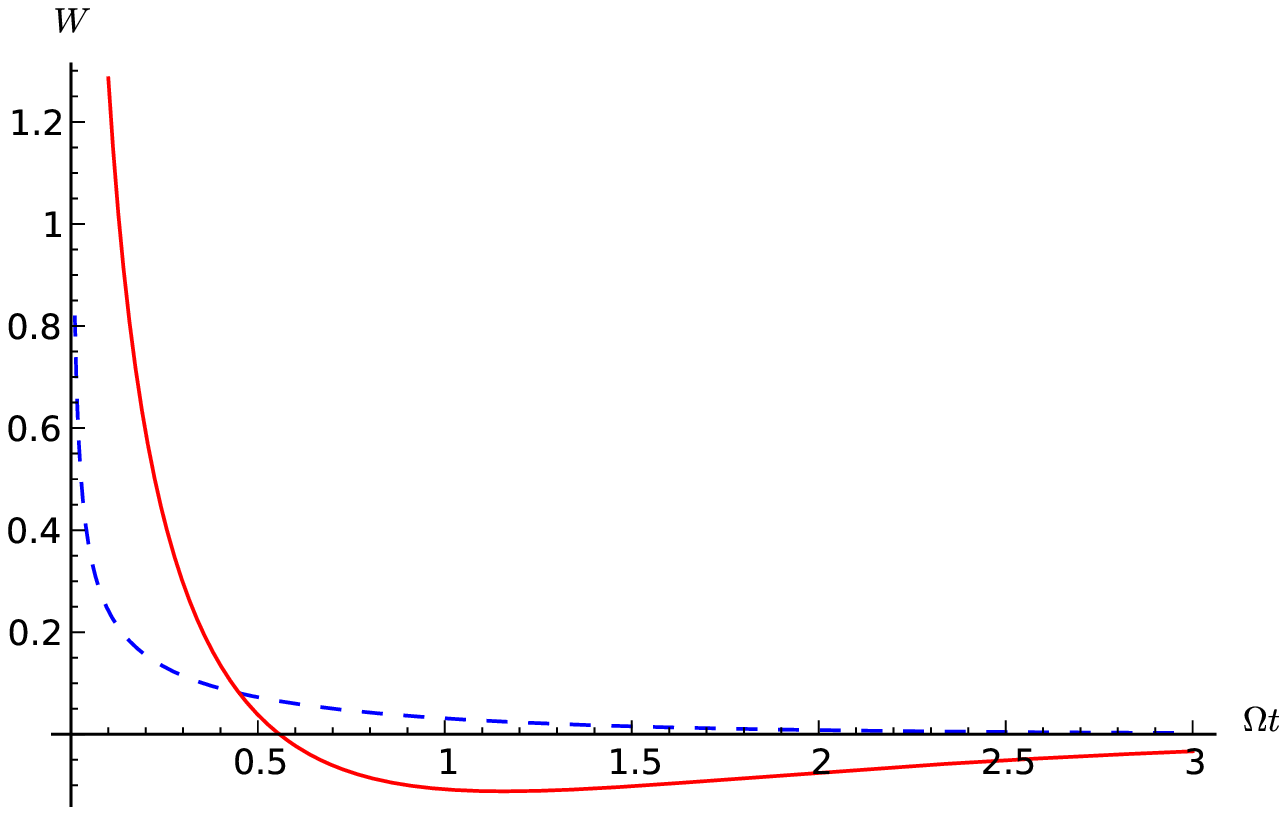}
\caption[k1] {[Color online] Plot of the Weighting function $W$ (eq. (\ref{Lap2}))(full line) and the Vardy-Brown weighting function $W_{VB}$ (eq. (\ref{VB}))(dashed), both divided by $\left(f/4\right){\mathbf{Re}}\:R\omega $, as functions of $\Omega t$}
\label{f1}
\end{figure}

\begin{figure}[htp]
\centering
\includegraphics[scale=.8]{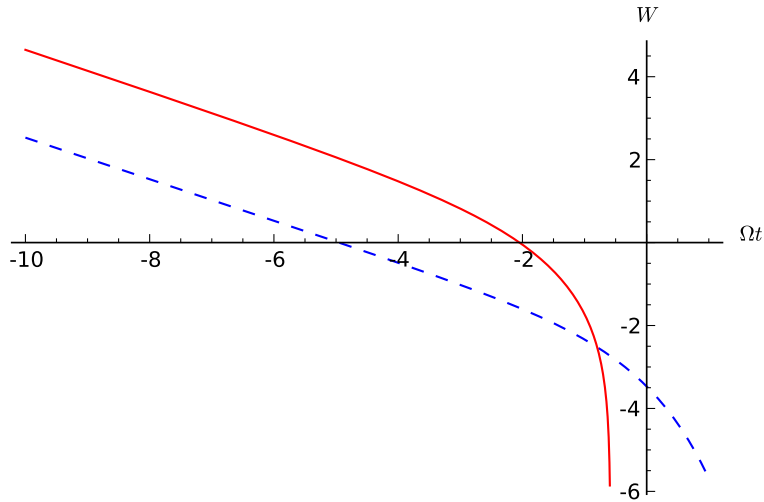}
\caption[k1] {[Color online] Log-Log plot of the Weighting function $W$ (eq. (\ref{Lap2}))(full line) and the Vardy-Brown weighting function $W_{VB}$ (eq. (\ref{VB}))(dashed), both divided by $\left(f/4\right){\mathbf{Re}}\:R\omega $, as functions of $\Omega t$}
\label{f2}
\end{figure}
\section{Final remarks}
In this paper we have shown a simple generalization of  the momentum transfer model to time-independent inhomogeneous flows, which yields a dynamic friction similar to the Brunone model \cite{Bru00}.

Matching the Vardy and Brown model \cite{VarBro03,VarBro04} proves to be a harder task. In this case we must appeal to a phenomenological approach to obtain a minimally acceptable MTM model

Although we have made several ad hoc choices in order to achieve this match, we believe there are at least two senses in which these results are meaningful.  

First, if it is conceded that a weakly time dependent mean flow produces a spectrum of turbulent fluctuations as depicted by the stochastic equation eq. (\ref{dynlinear}) with the noise self correlation given by eqs. (\ref{dynselfb}) and (\ref{tildeU}), then this may be used as a benchmark for more fundamental approaches such as those in \cite{EdMcC72} and \cite{Cal09b}.

Second, while the model is built to match Vardy and Brown's weighting function for a weakly time-dependent flow, it is clearly superior to it when space dependence becomes an issue, since it has the Brunone model built in. This synthesis of the IAM and CM expressions for unsteady friction is a legitimate prediction of the momentum transfer model, which it ought to be possible to contrast against experimental data.

We continue our research in these two promising directions.

\section*{Acknowledgements}
This work is supported by Universidad de Buenos Aires, CONICET and ANPCyT

We thank Fernando Minotti for discussions.

\end{document}